\documentclass[twocolumn,showpacs,amsmath,amssymb,prb]{revtex4}% Physical Review B
\usepackage{graphicx}% Include figure files
\usepackage{dcolumn}% Align table columns on decimal point
\usepackage{bm}% bold math
\usepackage{psfrag}
\usepackage{verbatim}

\usepackage{color}
\newcommand{\aver}[1]{\langle #1 \rangle}
\newcommand{\el}[2]{^{#2}\mathrm{#1}}
\newcommand{\unit}[1]{\ensuremath{\, \mathrm{#1}}}

\def\up{\uparrow}
\def\dn{\downarrow}

\begin{document}

\title{Spin exchange interaction with tunable range between graphene quantum dots}

\author{Matthias Braun\footnote{present address: AREVA NP GmbH, Erlangen, Germany.}}
\author{P.\ R.\ Struck}
\author{Guido Burkard}
\affiliation{Institute of Theoretical Physics C, RWTH Aachen University, D-52056 Aachen, Germany}
\affiliation{Department of Physics, University of Konstanz, D-78457 Konstanz, Germany}

\date{\today}

\begin{abstract}
We study the spin exchange between two electrons localized in separate quantum dots in graphene.
The electronic states in the conduction band are coupled indirectly by tunneling to a 
common continuum of delocalized states in the valence band. 
As a model, we use a two-impurity Anderson Hamiltonian which we subsequently transform into
an effective spin Hamiltonian by way of a two-stage Schrieffer-Wolff transformation.
We then compare our result to that from a Coqblin-Schrieffer approach as well as to fourth order
perturbation theory. 
\end{abstract}


\maketitle

\section{Introduction}
\label{sec:introduction}

Spins in quantum dots (QDs) are under intense investigation as a possible realization of a quantum 
bits (qubits).\cite{Loss1998} 
Among the currently most advanced solid-state structures are top-gate patterned two-dimensional electron gases in GaAs heterostructures.\cite{Hanson2007} However, in this host material hyperfine interaction between the spin of the electron and that of the atomic nuclei of the host material leads to relatively short coherence times.
A promising way to circumvent this problem is the use of carbon as a host material
for spin qubits.
Natural carbon comprises 99\% of the carbon isotope $\el{C}{12}$ which has no nuclear spin. 
This gives carbon based devices the advantage that decoherence due to hyperfine interaction is suppressed by the small abundance of nuclear spins. 
Carbon is also a relatively light element, therefore spin-orbit coupling is expected to be weaker than in GaAs. 
One can expect a significant improvement of spin coherence times in carbon based structures.
Graphene, a truly two-dimensional carbon-based crystal 
\cite{Novoselov2004,Novoselov2005,Zhang2005,Geim2007,
Katsnelson2007} seems to be an ideal host material 
for spin qubits\cite{Trauzettel2007,Tombros2007}.
It naturally created a perfect confinement of electrons in one dimension. 
Moreover, in contrast to carbon nanotubes,\cite{Sahoo2005,Shenoy2005} the ability to lithographically pattern graphene allows for a deterministic device preparation, which is necessary for scalability.\cite{Loss1998}
Graphene has a very interesting electronic structure with a gapless and linear dispersion around the Fermi energy.  Furthermore, the electronic eigenstates carry an additional internal degree of freedom, dubbed pseudospin, which is always aligned with the direction of the momentum.  These properties imitate the behavior of relativistic chiral massless Dirac particles.\cite{Zhang2005,Novoselov2005,Katsnelson2006} These relativistic-like properties lead to the phenomenon of Klein tunneling,\cite{Klein1929,Katsnelson2006} which actually prohibits any electrostatical confinement of electrons, i.e. prohibits the formation of quantum dots.

\begin{figure}[t!]
\includegraphics[width=0.95\columnwidth]{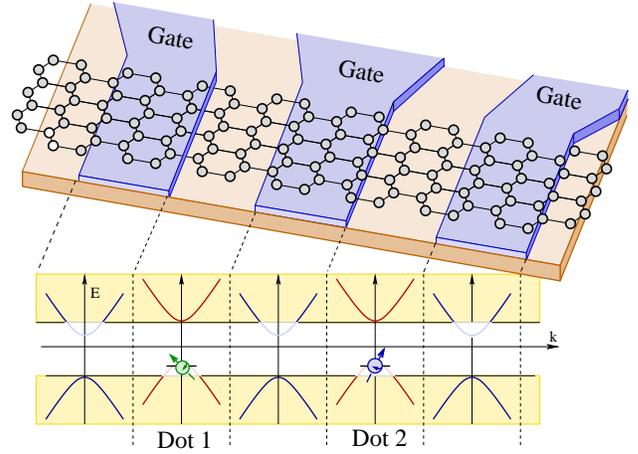}
\caption{\label{fig:ribbon}
One-dimensional quantum dot array on an armchair 
graphene nanoribbon (drawing not to scale). 
Due to the ribbon structure, the dispersion relation of graphene can exhibit a gap, which scales inversely with the ribbon width. This gap allows for electrostatic confinement of electrons in quantum dots. As the bandgap is small compared to regular semiconductors, the spin exchange mechanism between the quantum dots is not dominated by RKKY-type processes alone, and superexchange processes contribute significantly.}
\end{figure}

Among the most promising ideas to overcome Klein tunneling is to use graphene nanoribbons or constrictions instead of extended graphene as host material,\cite{Trauzettel2007,Stampfer2008,Silvestrov2007} see Fig.~\ref{fig:ribbon}. 
In clean graphene nanoribbons with armchair boundaries, the additional confinement can lead to the opening of a small energy gap at the Fermi energy.\cite{Brey2006}  The size of the gap is indirectly proportional to the width of the ribbon.  In the presence of such a gap, the pseudo-relativistic behavior of the charge carriers is lost, and the material resembles a regular gaped semiconductor, enabling electrostatic confinement.\cite{Trauzettel2007}  Interestingly, experiments observe the formation of a gap irrespective of the boundary condition.\cite{Han2007}  As a side effect, the sharp edges also lift the valley degeneracy in bulk graphene, which could suppress the Heisenberg spin interaction between the quantum dots.\cite{Trauzettel2007} 

Following Ref.~\onlinecite{Trauzettel2007}, we consider a system as shown in the upper part of Fig.~\ref{fig:ribbon} where several electric gates are placed on top of an armchair nanoribbon.  
By applying a gate voltage, the dispersion relation of the material below the gates can be shifted in energy, see the lower part of Fig.~\ref{fig:ribbon}. If at a certain energy extended states could exist in one section, but not in the neighboring ribbon sections, additional size quantization along the ribbon leads to single localized states. In the energy interval above and below the gap, the states are extended and form a continuum. With a suitable adjustment of the chemical potential, the localized states can be filled with one electron each, forming a one-dimensional array of qubits.  In the following, we calculate the Heisenberg spin interaction between two such spin qubits. 
%The coupling is assumed to be dominated by virtual tunnel processes over the (1-dimensional) continuum of states.\cite{Durganandini2006} 

The RKKY-interaction \cite{RKKY} between localized magnetic moments is well discussed for extended graphene, see Refs.~\onlinecite{Shung1986,Bastard1979}, and after the experimental discovery of graphene revisited in Refs.~\onlinecite{Ando2006,Wunsch2006,Saremi2007,Hwang2008,Dugaev2006}. 
Here, we study a graphene nanoribbon where a gap opens at the Fermi energy. Due to this gap, the spin exchange problem we are interested in resembles less the one in extended graphene but more the case in ordinary semiconductors in reduced dimensionalities, \cite{Aristov1997,Litvinov1998,Balcerzak2007} like quantum wells \cite{Bak2001} or quantum wires. \cite{Durganandini2006} 
However, compared to conventional semiconductors, the band gap can be unusually small, on the order of $1-100 \unit{meV}$. Therefore, by applying gate voltages, one can realize an arbitrary alignment of the quantum dot energy level relative to the valence and conduction band. Due to the proximity of the band edges of valence and/or conduction band, the band edges must be taken into account.\cite{Ziener2004}

\section{Transformation of the Hamiltonian}
\label{sec:technique}

\subsection{Model}
\label{sec:model}
We model the system as two Anderson impurities, which are both in 
contact with a common energy band. The Hamiltonian for our model is
\begin{eqnarray}
	\label{dots}
	H &=& H_0+H_T, \\
	H_0 &=& \sum_{i=1,2} \left( \sum_{\sigma=\uparrow,\downarrow} 
	\varepsilon_i a^{\dag}_{i\sigma}a_{i\sigma}
	+ U_i a^{\dag}_{i\up}a_{i\up}a^{\dag}_{i\dn}a_{i\dn} \right) \nonumber \\
	& & + \sum_{k\sigma} \varepsilon_{k} c^{\dag}_{k\sigma} c_{k\sigma}, \\
	H_T &=& \sum_{ik\sigma} \left(t_{ik}(\bm r_i) c^\dag_{k\sigma} a^{}_{i\sigma} 
	+ t_{ik}^{\star}(\bm r_i) a^{\dag}_{i\sigma} c_{k\sigma} \right).  
\end{eqnarray}
The first part of $H_0$ describes the two independent quantum dots $i=1,2$, containing one electronic level each. 
The fermionic operators $a^{}_{i\sigma}$ and $a^\dag_{i\sigma}$ create and annihilate electrons on dot $i$ with spin $\sigma$. 
Due to the low electrostatic capacity of a quantum dot, double occupation of one individual dot is associated with a substantial charging energy $U_i$. 
For quantum computation applications one is interested in a  parameter regime where the quantum dots are occupied with one electron each to form a spin 1/2 qubit.

The second part of $H_0$ models the contacting continuum of states  as a large reservoir of noninteracting electrons. 
Here $c_{k\sigma},c^{\dag}_{k\sigma}$  denote the annihilation and creation operators for electrons in the continuum with (orbital) quantum numbers $k$ and spin $\sigma$. 
The continuum is assumed to be unpolarized and  at zero temperature (filled valence band). 
The tunneling Hamiltonian $H_T$ describes spin-conserving tunneling between the two dots and the continuum. 
The tunnel amplitudes $t_{ik}(\bm r_i)$ depend on the dot and the continuum quantum numbers $i$ nd $k$, as well as on the position ${\bm r}_i$ of the quantum dots. However, the exact form of $t_{ik}(\bm r_i)$ depends on the system under consideration.

The two-impurity Anderson model and the strongly related Anderson lattice model is extensively discussed in the literature in the context of rare earth compounds,\cite{Bauer1991,Schlottmann2000,Proetto1981,Proetto1982,Hewson1977,Mohanty1981,Alascio1973, BookMahan,Coleman2002} dilute magnetic semiconductors,\cite{Singh2003} high temperature superconductors.\cite{Zaanen1988,Eid1995,Ihle1990} Numerous theoretical techniques were used to study these models, including  variational wave functions,\cite{Andreani2003,Simonin} equations of motion,\cite{Mohanty1981,Alascio1973} finite \cite{Grewe1981,Kuramoto1981,Proetto1982,Hewson1977,Zaanen1988} and infinite perturbation expansions,\cite{Bickers1987} Bethe-ansatz studies,\cite{Bazhanov2003} higher-order Schrieffer-Wolff transformations,\cite{Kolley1997,Kolley1998,Kolley1992,Tien1994, Ong2006,Ihle1990,Eid1995} or, most common, a mixture of a first-order Schrieffer-Wolff transformation and perturbation theory,\cite{Coqblin1969, Cornut1972,Yang2005,Schlottmann2000, Siemann1980, Singh2003, Tamura2004,Castro1995,Proetto1981} originally proposed by Coqblin and Schrieffer in Ref.~\onlinecite{Coqblin1969}.

\subsection{Schrieffer-Wolff Transformation}

Following Refs.~\onlinecite{Kolley1992,Tien1994,Kolley1997,Kolley1998} we use a two-stage or nested Schrieffer-Wolff transformation to derive an effective spin Hamiltonian. In contrast to previous works we do not assume equal energy levels in the two Anderson impurities, as the confinement of the quantum dots can be modified individually. By keeping track of the dot indices it is also possible to identify different physical processes in the final result, and enables us to compare our result to higher-order perturbation theory.

The Schrieffer-Wolff transformation\cite{Schrieffer1966, Cornut1972, BookFlensberg,BookHewson,BookMahan} is based on a canonical transformation of the Hamiltonian, $H^{(1)}=e^{iS}\,H\,e^{-iS}= H+ [iS,H]+\frac{1}{2}[iS,[iS,H]]+\ldots$. 
The division of the Hamiltonian $H$ into a free Hamiltonian $H_0$ and 
a small perturbation $H_T$, allows us to choose a transformation $S_1$, fullfilling the relation $[iS_1,H_0]=-H_T$ and  leading us to the effective Hamiltonian $H^{(1)}= H_0+\frac{1}{2}[iS_1,H_T]+\frac{1}{3}[iS_1,[iS_1,H_T]]+\frac{1}{8}[iS_1,[iS_1,[iS_1,H_T]]]+...$ where the lowest-order tunneling term is canceled (note that the coefficient has the general form $1/n!-1/(n+1)! = 1/[(n-1)!(n+1)!]$). 
Since $S$ also has to be of first order in the tunneling amplitudes,  $S_1\propto H_T$, the interaction now appears (at least) in second order. 

By a subsequent Schrieffer-Wolff transformation with the generator $S_2$ fulfilling $[iS_2,H_0]=-\frac{1}{2}[iS_1,H_T]$, also the second order interaction term can be removed.  Note that now, $S_2\propto H_T^2$. 
At the end we project the resulting Hamiltonian on the subspace where both quantum dots are occupied by one electron. As all odd-order interactions do not conserve the occupation numbers of the quantum dot, they can be neglected, as they will be projected out at the end of the calculation.  Combining both steps, we arrive at the effective Hamiltonian 
\begin{eqnarray}\label{eq:hamiltonian2}
H^{(2)}=H_0+\frac{1}{4}[iS_2,[iS_1,H_T]]+\frac{1}{8}[iS_1,[iS_1,[iS_1,H_T]]],\quad
\end{eqnarray}
where corrections in sixth and higher orders in the tunneling amplitudes have been neglected. 
After projecting out the continuum degrees of freedom, and in addition to unimportant level renormalizations, which we do not discuss, we find a Heisenberg-like interaction $J \bm S_1 \cdot \bm S_2$, which couples the two quantum dot spins $\bm S_i= \sum_{\alpha\beta} a^\dag_{i\alpha}\,{\bm \sigma}_{\alpha \beta}\,a_{i\beta}$, consistent with Refs.~\onlinecite{Kolley1997,Kolley1998}. 
A detailed calculation is presented in the Appendix. After non-trivial regrouping of terms \cite{note1}, one can separate the spin-interaction into parts originating from different virtual tunneling processes defined by their intermediate virtual quantum state with the explicite shape:
\begin{widetext}
\begin{eqnarray}
	J &=& 2 \sum_{k,q}t^{\star}_{1,k} t_{2,k} t_{1,q}t^{\star}_{2,q} \,
	e^{i({\bm k}-{\bm q})\cdot({\bm R}_1-{\bm R}_2)}\,(\,J_1+J_2+J_3+J_4\,) \label{r0}, \\
	J_1 &=& \left(\frac{1}{\varepsilon_k-\varepsilon_1}-\frac{1}{\varepsilon_q-\varepsilon_1-U_1}\right) 
	\frac{n_k-n_q}{\varepsilon_k-\varepsilon_q}
	\left(\frac{1}{\varepsilon_q-\varepsilon_2}-\frac{1}{\varepsilon_k-\varepsilon_2-U_2}\right) \label{r1}, \\
	J_2 &=& \left(\frac{1}{\varepsilon_k-\varepsilon_1}+\frac{1}{\varepsilon_q-\varepsilon_2}\right) 
	\frac{1-n_q}{\varepsilon_k+\varepsilon_q-\varepsilon_1-\varepsilon_2}
	\left(\frac{1}{\varepsilon_q-\varepsilon_1}+\frac{1}{\varepsilon_k-\varepsilon_2}\right) \label{r2}, \\
	J_3 &=& \left(\frac{1}{\varepsilon_1+U_1-\varepsilon_k}+\frac{1}{\varepsilon_2+U_2-\varepsilon_q}\right) 
	\frac{+n_q}{\varepsilon_1+U_1+\varepsilon_2+U_2-\varepsilon_k-\varepsilon_q}
	\left(\frac{1}{\varepsilon_1+U_1-\varepsilon_q}+\frac{1}{\varepsilon_2+U_2-\varepsilon_k}\right) \label{r3}, \\
	J_4 &=& \left(\frac{1}{\varepsilon_k-\varepsilon_1}+\frac{1}{\varepsilon_k-\varepsilon_2-U_2}\right) 
	\frac{-n_q}{\varepsilon_2+U_2-\varepsilon_1} 
	\left(\frac{1}{\varepsilon_q-\varepsilon_1}-\frac{1}{\varepsilon_q-\varepsilon_2-U_2}\right)
	+\frac{1}{\varepsilon_k-\varepsilon_1} \frac{+1}{\varepsilon_2+U_2-\varepsilon_1} 
	\frac{1}{\varepsilon_q-\varepsilon_1}
	+(1\leftrightarrow 2)\,.\label{r4}\qquad 
\end{eqnarray}
\end{widetext}

\begin{figure}[t!]
\includegraphics[width=0.85\columnwidth]{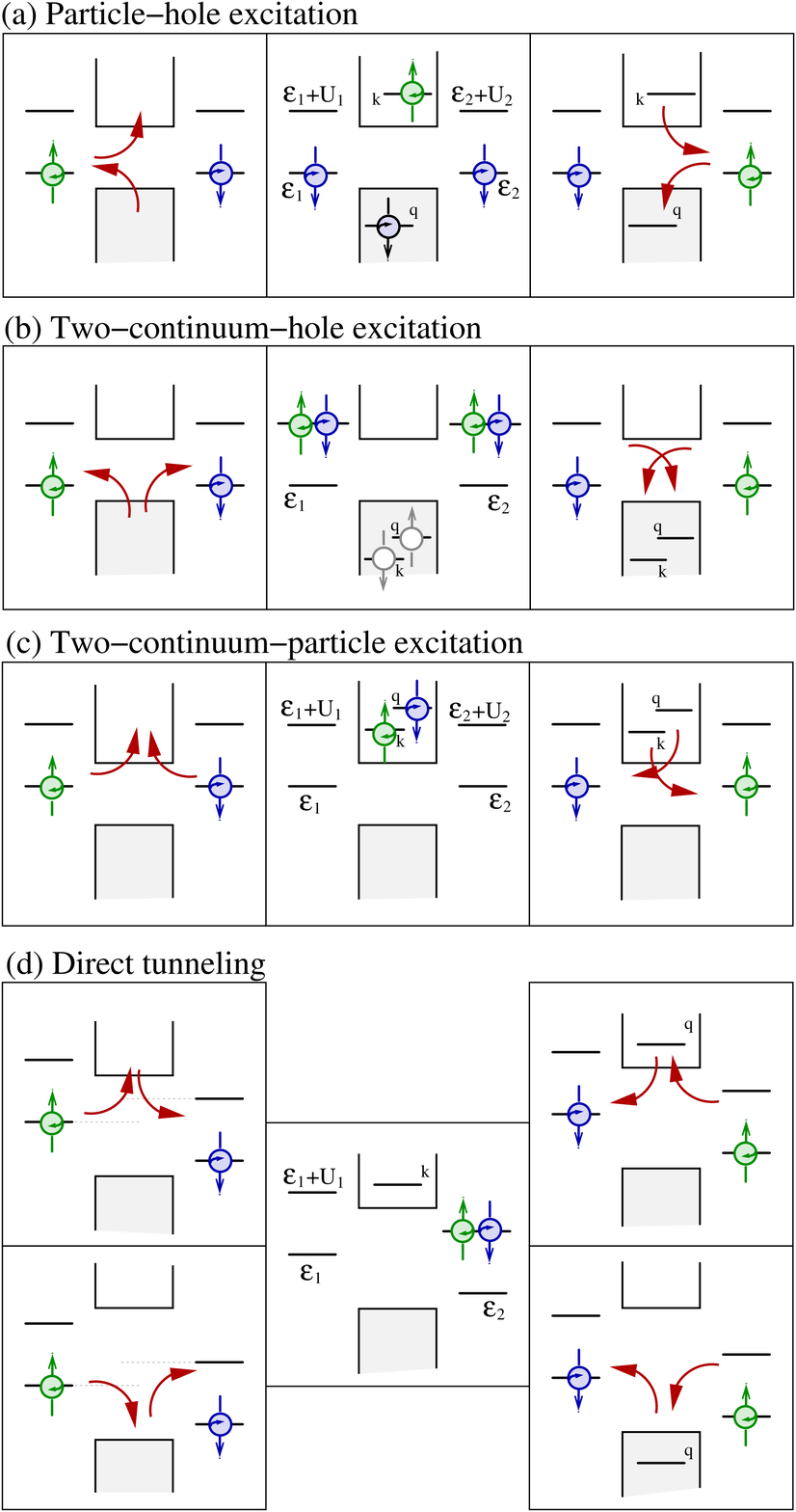}
\caption{\label{fig:processes}
Several virtual tunnel processes contribute to the spin-spin interaction between the dots. These processes can be classified by the intermediate state of the system. While particle-hole excitation (a) leads to an RKKY-like interaction, the processes (b)-(d) are usually summarized as superexchange.}
\end{figure}

The first term $J_1$ resembles an RKKY-interaction.\cite{RKKY} The interaction is mediated by a virtual particle-hole excitation in  the electron gas, see Fig.~\ref{fig:processes}(a). Therefore the energy of the intermediate excitation is given by $\varepsilon_q-\varepsilon_k$.
The second and third contributiosn to the spin-spin interaction originate from virtual two-particle(hole) excitations in the continuum Fermi sea, see Fig.~\ref{fig:processes}(b) and (c). 
Two electrons tunnel coherently from or to the quantum dots. Thus, as intermediate virtual states, the two quantum dots are both doubly occupied (empty). Afterwards, the electrons tunneling crosswise back, interchanging the spins of the quantum dots. This process leads to the interactions $J_2$ and $J_3$.

Finally, the last contribution $J_4$ is caused by the possibility of a direct tunneling of one dot electron to the other dot. The virtual intermediate state is therefore one double occupied dot and one empty dot. The tunneling can happen through filled as well as empty states in the electron gas, see Fig.~\ref{fig:processes}(d)

\subsection{Relation to Coqblin-Schrieffer Model}
\label{sec:derivation}
In Ref.~\onlinecite{Coqblin1969} Coqblin and Schrieffer presented their widely used approach 
\cite{Schlottmann2000,Coleman1986,Bazhanov2003,Yang2005} to the two-impurity Anderson model. They performed a single Schrieffer-Wolff transformation, and project the resulting Hamiltonian of the single-occupied impurities. With this, they transform the two individual Anderson models into two individual s-d models or Kondo impurities. The spin of one quantum dot ${\bm S}_i$ is coupled to the bath electron spins by 
\begin{eqnarray}\label{eq:Kondo}
H_{\rm Kondo}=\sum_{kq,\sigma\sigma^\prime}J^i_{kq} {\bm S}_i\cdot\, c^\dag_{k\sigma}{\bm \sigma}_{\sigma\sigma^\prime}c^{}_{q\sigma^\prime}\,,
\end{eqnarray}
see Appendix A for details. By treating that Hamiltonian in second order perturbation theory, they compute a RKKY-like spin-spin interaction\cite{RKKY} of the form 
\begin{eqnarray}\label{eq:KondoRKKY}
\sum_{kq}J^1_{kq}J^2_{qk} \frac{n_k-n_q}{\varepsilon_k-\varepsilon_q}{\bm S}_1\cdot {\bm S}_2 .
\end{eqnarray}
Even though Eq.~(\ref{eq:KondoRKKY}) captures the basic features of Eq.~($\ref{r1}$), it is an expansion inconsistent in the order of the tunnel amplitudes. First, the initial Schrieffer-Wolff transformation generates not only terms of second order, but also the term $[iS_1,[iS_1,[iS_1,H_T]$ which contributes in fourth order,\cite{Castro1995} see Eq.~(\ref{eq:hamiltonian2}). Actually contributions from this higher-order term cancel several parts in Eq.~(\ref{eq:KondoRKKY}), leading to Eq.~(\ref{r1}). By truncating the transformation at second order, these contributions to the spin interaction are lost. 

Second, the initial Schrieffer-Wolff transformation does not only generate the Kondo-Hamiltonian, but also terms in second order of the form $a_1^\dag a_2$, which describe direct tunneling between the two quantum dots.\cite{BookMahan} 
In a subsequent second order perturbation theory, these terms also lead to a inter-dot spin interaction. 
In Ref.~\onlinecite{Coqblin1969} these interactions are neglected due to the premature projection of the result of the Schrieffer-Wolff transformation on the single occupied dot subspace.

Interestingly, in the limit of energy levels far away from the Fermi energy, i.e., when one assumes that the spin coupling $J^i_{kq}$ approaches a constant, the Coqblin-Schrieffer approach generates the correct result. However, nowadays the Anderson model is extensively used to describe quantum dots.\cite{Durganandini2006,Tamura2004,Yang2005} In contrast to rare earth compounds or true atomic systems, the typical energy scale of the quantum dot level spectrum is several orders of magnitude smaller. Therefore in these artificial systems the application of the Coqblin-Schrieffer model needs to be handled with care.

\subsection{Relation to fourth order perturbation theory}
\label{sec:perturbationtheory}
Starting from the two-impurity Anderson model, one can also derive a fourth order dot spin-spin interaction by perturbation theory,\cite{Zhang1986,Proetto1981,Proetto1982, Grewe1981, Kuramoto1981, Hewson1977} with or without diagrammatical help. The perturbation approach nearly reproduces our results Eq.~(\ref{r1})-(\ref{r4}) with one exception: the structure of the Fermi functions. Via perturbation theory one would expect for example that the two-lead-particle excitation, see Fig.~\ref{fig:processes}(c), only happens if the two electron gas states $k$ and $q$ are empty, therefore the spin coupling $J_2$ should be proportional to $(1-n_k)(1-n_q)$. In contrast, the contribution from a  Schrieffer-Wolff transformation is proportional to $(1-n_q)$, 
and independent of $n_k$.
By counting the operator commutators, one can directly determine that the spin-spin coupling derived by a two-stage Schrieffer-Wolff transformation can not generate terms, which contain four lead operators, which would be necessary for a product term like $n_k  n_q$. 
The reason for this discrepancy between fourth order perturbation theory and Schrieffer-Wolff transformation lies in the procedure of integrating out the lead degrees of freedom, i.e., by the replacement of the thermal average of lead operators $\aver{c^{\dag}_k c_k}_{\rm{th}}$ by the Fermi function $n_k$. 
In the case of the perturbation theory, the operators $c^{\dag}_k,c_k$ refer to the bare unperturbed electronic states of the lead, i.e., one assumes, that the lead is actually not perturbed by tunneling.  After the Schrieffer-Wolff transformation, the lead operators refer to new lead states, which are hybridized with the localized dot states. By performing the thermal average, one therefore assumes that these new hybridized lead states are in thermal equilibrium, not the bare lead states. 
Therefore, it is not surprising that the results of a Schrieffer-Wolff transformation and perturbation theory differ. 
However, it is surprising that one can express the result of the Schrieffer-Wolff transformation in the same functional form one would expect from perturbation theory, except for the Fermi functions. Only due to this structure of terms  one actually can, in the spirit of Feynman diagrams, assign virtual processes as shown in Fig.~\ref{fig:processes}. For this reason, the grouping of terms in Eq.~(\ref{r1})-(\ref{r4}) is physically plausible but to some extent arbitrary.
It has already been observed by Ruderman and Kittel\cite{RKKY} that by assuming certain symmetries of the tunnel amplitudes, the Pauli exclusion principle can actually become in part unimportant.

\section{Application to graphene nanoribbon quantum dots}
\label{sec:application}
Up to now, the computed result in Eq.~(\ref{r1}-\ref{r4}) is general for the spin coupling of two qubits by a common continuum of states labeled by the indeces $k$ and $q$. In the following, we specify this continuum to the electronic structure of a graphene nanoribbon aligned along the  $y$-direction.

\subsection{Band structure}
\label{subsec:dispersion}
Bulk graphene has two independent Fermi points at the momenta $\bm K$ and $\bm K^\prime$ in reciprocal space, generating the valley degeneracy. 
Due to the armchair boundary conditions of the ribbon, the propagating wave states with momentum $\bm K+\bm k$ and $\bm K^\prime+\bm k$ are coupled.\cite{Brey2006,Brey2007} 
The confinement in x-direction leads to a further quantization of the transverse wave vector $k_x\equiv k_n=(n\pm1/3)\pi/W$ with $W$ denoting the ribbon's width and $n\in \mathbb{N}$. 
Therefore the continuum states can be characterized by the subband index $n$ and the momentum component $k_y \equiv k$ along the ribbon. Close to the Fermi energy, the dispersion relations becomes
\begin{eqnarray}\label{eq:dispersion}
\varepsilon_{k,n}=\hbar v_F\sqrt{k^2+k_n^2},
\end{eqnarray}
with the Fermi velocity $v_F$ of graphene. This dispersion resembles the dispersion of a massive relativistic particle.

The transverse confinement determines the energy gap $2 \varepsilon_g= 2\hbar v_F k_0$, which scales inversely with the ribbon width.\cite{Han2007}  
Due to this gap, electrons can be confined by electrostatic gates, in analogy to conventional semiconductors.\cite{Trauzettel2007} 
We assume  the applied electric potential to be independent of the x-coordinate (see Fig.~\ref{fig:ribbon}), therefore the band index $n$ is conserved. 
Therfore we only need to consider the continuum subband with the same band index as the bound state(s). 
Even if this symmetry is broken, the generalization to multi-subbands is straight forward.\cite{Shenoy2005}
As a further simplification, we assume that by applying electrostatic gates, the dispersion relation of the extended states is still described by Eq.~(\ref{eq:dispersion}).

\subsection{Tunnel amplitudes}
\label{subsec:tunnelamplitudes}

The spin exchange is proportional to the product of the four tunnel amplitudes
$t^\star_{1,k}(\bm r_1) t^{}_{2,k}(\bm r_2) t^{}_{1,q}(\bm r_1) t^\star_{2,q}(\bm r_2)$. 
In analogy to most cases studied in literature, we assume that the amplitude of the overlap of the bound states and the extended states does not explicitely depend on the momentum $k$ of the extended states. This assumption is valid if the wave function of the bound state is localized on a length scale smaller than the wave length of the extended state. In this case, one can approximate the localized wave function as a delta-function. However, in quantum dots in semiconductors in general, and in particular in the vicinity of a band edge, this assumption may  not be valid. As $k$-independent tunnel amplitudes lead to a shorter spin exchange range, the spin exchange range derived in the following can be seen only as lower bound.  Although the magnitude of the tunnel amplitude does not depend on $k$ within this approximation, the fact that the two quantum dots are separated in space gives rise to a relative phase. While in ordinary isotropic metals this phase is simply given by $e^{i \bm k\cdot \bm r_i}$, in graphene the valley degeneracy has to be taken into account. In nanoribbons, the energy eigenstates are phase-locked superposition of states of both valleys. Therefore, the overlap of the wave functions\cite{Trauzettel2007,Brey2007} leads to a tunnel amplitude of the form
\begin{eqnarray}\label{eq:tunnelamplitude}
t^{}_{1,k}(\bm r_1)= t^{}_{1} e^{-i \bm k\cdot \bm r_1}\frac{e^{-i \bm K\cdot \bm r_1} +e^{-i \bm K^\prime\cdot \bm r_1}}{\sqrt{2}}.
\end{eqnarray}
The spin coupling therefore will always contain a contribution which oscillates on inter-atomic distances, and one contribution, which varies on the length scale of the envelope wave function. As the quantum dots under consideration are not spatially defined with lattice site precision, we expect that the oscillating contribution to the spin exchange will average out.

\subsection{Spin-exchange range}
\label{subsec:range}

\begin{figure}[t!]
\includegraphics[width=0.45\columnwidth]{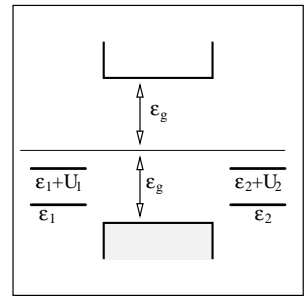}
\caption{\label{fig:situations}
If the quantum dot level lies close to the valence band, and the charging energy is smaller than the band gap, direct tunneling processes will dominate the spin exchange between the quantum dots.}
\end{figure}

Which of the virtual tunnel process (see Fig.~\ref{fig:processes}) dominates the spin exchange between two quantum dots depends on the alignment of the dot energy levels, band gap, and edges. 
Roughly speaking, the virtual process requiring the lowest excitation energy will dominate.
As we assume the graphene nanoribbon to be nearly undoped, the Fermi energy of the system lies within the band gap. 
Therefore the valence band is entirely filled, and the conduction band is empty. 
The RKKY-like exchange interaction via a particle-hole excitation in the continuum will be suppressed by the band gap $2 \varepsilon_g$, and superexchange processes will dominate. In the following, we will discuss the scenario shown in Fig.~\ref{fig:situations}, where the quantum dot level lies close to the valence band, and the charging energy is smaller than the band gap.

In this level alignment, direct tunneling processes via the valence band dominate the spin exchange. The resulting integrals can be computed by Cauchy's integral formula. The exchange due to direct tunneling turns out to be 
\begin{eqnarray}
	J &=& -\frac{|t_1|^2 |t_2|^2}{4 \Delta E^2} \frac{1}{\varepsilon_2+U_2-\varepsilon_1}
	\frac{(\varepsilon_2+U_2)^2}{\varepsilon_g^2-(\varepsilon_2-U_2)^2} \nonumber \\
	& & \times e^{-2 \frac{\sqrt{\varepsilon_g^2-(\varepsilon_2+U_2)^2}}{\hbar v_F}\Delta r}
	+(1\leftrightarrow 2) \label{J-result}
\end{eqnarray}
with the quantum dot distance $\Delta r=|\bm r_1-\bm r_2|$. The energy  $\Delta E=\hbar v_F/L$ is the energy splitting of the continuum states, with $L$ being the length of the ribbon. As the tunnel amplitudes $t_i$ decrease with the real space particle density of the band states with $1/\sqrt{L}$, the strength of the spin exchange is independent of the overall length of the ribbon.

The strength of the spin coupling driven by direct tunneling diverges within $4^{th}$ order, if the single occupied state of one dot becomes resonant with the double occupied state of the other dot. %
The range $\lambda =  \hbar v_F / 2\sqrt{\varepsilon_g^2-(\varepsilon_2+U_2)^2}$ 
of the coupling Eq.~(\ref{J-result}) on the other side is controlled by the energy separation between the double occupied state of the one dot, and the valence band edge at energy $-\varepsilon_g$. 
If one assumes, that the single occupied states of the two dots are close to the band edge ($\varepsilon_g\approx \varepsilon_i$), the exchange range scales as $\hbar v_F/\sqrt{8 \varepsilon_g U_i}$. 
For a graphene nanoribbon with width of $50 \unit{nm}$ and a quantum dot with a charging energy of $4 \unit{meV}$,\cite{Stampfer2008} this length is of order $50 \unit{nm}$, i.e. comparable to the quantum dot length.

If one assumes that, in analogy to a capacitor,  the charging energy of a quantum dot scales inversely with its area,\cite{Stampfer2008} then the range of the spin exchange interaction scales linearly with the width of the nanoribbon.

%In the situation shown in Fig.~\ref{fig:situations}(b), when the charging energy of a quantum dot is large compared to the band gap, the longest spin exchange is reached, if the dot levels approach the empty states of the valence band. As the application as spin qubits require the occupation of the quantum dot by one electron, the Fermi energy must lie between the quantum dot levels and the band edge. In this regime the spin exchange range is approximatively given by $\hbar v_F/\sqrt{8 \varepsilon_g \delta e}$, whereby $ \delta e$ the energy difference between the quantum dot levels and the band edge labels. Theoretically, this would give rise to an arbitrary  

\subsection{Further considerations}
\label{subsec:considerations}

The lower bound for the spin exchange range between quantum dots in a graphene nanoribbon is given by the ribbon width. For this result we considered only virtual tunnel processes via free continuum states. In addition, also direct tunneling between the quantum dot states can occur. If the quantum dot level approaches the band gap, the bound electron leaks further and further into the barrier due to the weakening of the confinement. Therefore it can happen that two neighboring quantum dot wave functions can acquire a non-vanishing overlap, and direct tunneling becomes possible. 
Direct tunneling is accompanied also by a spin-exchange interaction.\cite{Trauzettel2007}

Furthermore, we assumed so far that the graphene nanoribbon is infinitely long. This assumption is hidden in the approach to treat the continuum states with momentum $k$ independent on the state $-k$. However, if a finite ribbon length leads to a defined phase relation of the forward and backward propagating states, then one part of the tunnel amplitude will become entirely independent on the momentum $k$, and therefore on the distance $\Delta \bm r$.\cite{Trauzettel2007} (cf. also the discussion of section \ref{subsec:tunnelamplitudes}.) If this case, the range $\lambda$ of the spin exchange is not determined by the dephasing of the exchange contributions of different states within the Fermi sea, but only on the phase coherence length of the extended states.

\section{Conclusions}
\label{sec:conclusions}

In this paper, we discussed the spin exchange between localized states which are only indirectly coupled via a continuum of states.
Using a 2-stage Schrieffer-Wolff transformation, we transformed a 2-impurity Anderson model into an effective spin Hamiltonian. 
Based on our result, we discussed the validity of the Coqblin-Schrieffer approach to this problem. 
Furthermore, by a re-ordering the terms, we were able to directly compare the Schrieffer-Wolff result to  perturbation theory, and observe distinct differences that originate from different assumptions on the continuum of states.

As an application of the formalism developed here, we discussed the spin exchange interaction between electrostatically confined quantum dots in a graphene nanoribbon, as shown in Eq.~(\ref{J-result}). 
As a lower bound, we derive a range of this spin exchange of the order of the nanoribbon width.
However, the dot energies can be adjusted in such a way as to extend the exchange coupling to 
longer distances.

\section*{Acknowledgements}
We acknowledge funding from the DFG within FOR 912 and within SFB 767.

\appendix

\section{First Schrieffer-Wolff Transformation}
\label{app:first}
The generator of the first Schrieffer-Wolff transformation must fullfill the relation $[iS_1,H_0]=-H_T$. 
As the part of the Hamiltonian $H_0$ only contain number operators, i.e., is quadratic or quartic in the fermion operators, one can deduce, that $iS$ must have the same structure as the tunnel Hamiltonian $H_T$, up to a prefactor containing either constants or number operators. With such an ansatz, the generator of the first Schrieffer-Wolff transformation is found to be 
\begin{eqnarray}
&& iS_1= \\
&& \sum_{i,k,\sigma}\frac{t_{i,k}}{\varepsilon_k-\varepsilon_i-U_i n_{i\bar{\sigma}}}c^\dag_{k\sigma}a_{i\sigma}
-\frac{t^\star_{i,k}}{\varepsilon_k-\varepsilon_i-U_i n_{i\bar{\sigma}}}a^\dag_{i\sigma}c_{k\sigma}\qquad\nonumber \\
&& = \sum_{i,k,\sigma}t_{i,k} \left[ 
\frac{1-n_{i\bar{\sigma}}}{\varepsilon_k-\varepsilon_i}
+\frac{n_{i\bar{\sigma}}}{\varepsilon_k-\varepsilon_i-U_i n_{i\bar{\sigma}}}\right]
c^\dag_{k\sigma}a_{i\sigma}-{\rm h.c.}\nonumber
\end{eqnarray}
This transformation removes the interaction term of first order in the tunneling amplitude, but instead generates higher order interactions starting in second order of $t$. The new interaction Hamiltonian $H_T^{(1)}= \frac{1}{2}[iS_1,H_T]+O(t^3)$ becomes 
\begin{eqnarray}
	\label{eq:h1}
	H_T^{(1)} 
	&=& \sum_{i,kq,\sigma} J^{i}_{kq} {\bm S}_{i} \cdot {\bm s}_{kq} \\
	&+& \frac{-1}{2}\sum_{ij,k,\sigma} A^k_{ij} a^{\dag}_{i\sigma}a_{j\sigma} \nonumber \\
	&+& \mbox{dot double empty/filling terms} \nonumber \\
	&+& \mbox{spin independent lead scattering-terms} \nonumber
\end{eqnarray}
The first term resembles the Kondo model. The spin of the quantum dot ${\bm S}_i=\sum_{\sigma \sigma^\prime}a^\dag_{i\sigma}\, {\bm \sigma}^{}_{\sigma\sigma^\prime} \,a^{}_{i\sigma^\prime}$ couples to the band spin 
${\bm s}_{kq}=\sum_{\sigma \sigma^\prime}c^\dag_{k\sigma}\, {\bm \sigma}^{}_{\sigma\sigma^\prime} \,c^{}_{q\sigma^\prime}$. The coupling strength is given by 
$J^i_{kq}=t^\star_{i,q}t_{i,k} \left[
\frac{1}{\varepsilon_k-\varepsilon_i}-\frac{1}{\varepsilon_k-\varepsilon_1-U_1}+
\frac{1}{\varepsilon_q-\varepsilon_i}-\frac{1}{\varepsilon_q-\varepsilon_1-U_1}\right]$. The second term describes a direct tunneling of one quantum-dot electron to another dot with the effective coupling strength
$A^k_{ij}=t^\star_{i,k}t_{j,k}
\left[
\frac{1-n_{i\bar{\sigma}}}{\varepsilon_k-\varepsilon_i}
+\frac{n_{i\bar{\sigma}}}{\varepsilon_k-\varepsilon_i-U_i}
+\frac{1-n_{j\bar{\sigma}}}{\varepsilon_k-\varepsilon_j}
+\frac{n_{j\bar{\sigma}}}{\varepsilon_k-\varepsilon_j-U_j}
\right]$ 
with $\bar{\sigma}$ denoting the opposite spin orientation of $\sigma$. 
This term can also lead to a spin exchange in fourth order, so it is not negligible. The further parts of Eq.~(\ref{eq:h1}) include processes, which change the occupation of one quantum dot by two electrons, and spin-independent scattering of continuum electrons at one dot.

\section{Second Schrieffer-Wolff Transformation}
\label{app:second}
For the second Schrieffer-Wolff Transformation, the procedure is very similar. Using the ansatz for $iS_2$ that resembles the second order part of the interaction term, derived by the first transformation. However, as one is in the end interested in the fourth-order parts of the Hamiltonian, which couple two quantum dot spins and conserve the quantum dot occupation number, one only needs to consider the first two parts of Eq.~(\ref{eq:h1}). The generator for the transformation therefore can be written as $iS_2=iS_2^{(a)}+iS_2^{(b)}$ with
\begin{eqnarray}\label{eq:h2}
iS_2^{(a)}&=&\sum_{i,kq,\sigma} \frac{1}{\varepsilon_k-\varepsilon_q}J^i_{kq} {\bm S}_i\cdot {\bm s}_{kq}\\
iS_2^{(b)}&=&\frac{-1}{2}\sum_{ij,k,\sigma} 
\frac{1}{\varepsilon_i+U_i n_{i\bar{\sigma}}-\varepsilon_j-U_j n_{j\bar{\sigma}}}
A^k_{ij} a^\dag_{i\sigma}a_{j\sigma}\nonumber
\end{eqnarray}
The other second order terms finally drop out in the end, when the Hamiltonian is projected on the subspace of single occupied quantum dots. For the regrouping of terms in Eq.~(\ref{r0}-\ref{r4}), one needs the symmetry of the expressions under the replacement $1\leftrightarrow 2$ and $k\leftrightarrow q$.


\begin{thebibliography}{99}

%-----------------------------
%Introduction
%-----------------------------

%Quantum computation with quantum dots 
\bibitem{Loss1998}
D. Loss and D. P. DiVincenzo, Phys. Rev. A 57, 120 (1998).

%Spins in few-electron quantum dots 
\bibitem{Hanson2007}
R. Hanson, L. P. Kouwenhoven, J. R. Petta, S. Tarucha, and L. M. K. Vandersypen,
Rev.\ Mod.\ Phys.\ {\bf 79}, 1217 (2007).



%-----------------------------
%GRAPHENE
%-----------------------------

% Discovery of graphene
\bibitem{Novoselov2004} 
K.S. Novoselov, A.K. Geim, S.V. Morozov, D. Jiang, Y.
Zhang, S.V. Dubonos, I.V. Grigorieva, and A.A. Firsov, Science {\bf 306}, 666 (2004).

\bibitem{Novoselov2005}
K.S. Novoselov, A.K. Geim, S.V. Morozov, D. Jiang, M.I. Katsnelson,
I.V. Grigorieva, S.V. Dubonos, and A.A. Firsov, Nature {\bf 438}, 197 (2005).

\bibitem{Zhang2005} 
Y. Zhang, Y.-W. Tan, H.L. Stormer, and P. Kim, Nature {\bf 438}, 201 (2005).

\bibitem{Geim2007}
A.K. Geim and K.S. Novoselov, Nature Mater. {\bf 6},  183 (2007).

\bibitem{Katsnelson2007}
M.I. Katsnelson, Mater. Today {\bf 10}, 20 (2007).

% Spin qubits in graphene quantum dots
\bibitem{Trauzettel2007}
B. Trauzettel, D. V. Bulaev, D. Loss, and G. Burkard, Nature Physics 3, 192 (2007), cond-mat/0611252. 

%%Electronic spin transport and spin precession in single graphene layers at room temperature
\bibitem{Tombros2007}
N. Tombros, C. Jozsa, M. Popinciuc, H. T. Jonkman, B. J. van Wees, Nature \textbf{448}, 571-574 (2007).

%Electric field control of spin transport
\bibitem{Sahoo2005}
S. Sahoo, T. Kontos, J. Furer, C. Hoffmann, Matthias Gr\"aber, A. Cottet and C. Sch\"onenberger,
Nature Physics {\bf 1}, 99 (2005).

%RKKY interaction in single-walled nanotubes
\bibitem{Shenoy2005}
V. B. Shenoy,  Phys. Rev. B 71, 125431 (2005).

% electronical application
\bibitem{Katsnelson2006} 
M.I. Katsnelson, K.S. Novoselov, and A.K. Geim, Nature Phys. {\bf 2}, 620 (2006).

\bibitem{Klein1929}  
O. Klein, Z. Phys. {\bf 53} 157 (1929).

%Tunable Coulomb blockade in nanostructured graphene
\bibitem{Stampfer2008}
C. Stampfer, J. G\"uttinger, F. Molitor, D. Graf, T. Ihn, and K. Ensslin, Appl. Phys. Lett. {\bf 92}, 012102 (2008).

%Quantum Dots in graphene
\bibitem{Silvestrov2007}
P. G. Silvestrov and K. B. Efetov, Phys. Rev. Lett. {\bf 98}, 016802 (2007).

%Electronic states of graphene nanoribbons studied with the Dirac equation
\bibitem{Brey2006}
L. Brey and H. A. Fertig, Phys. Rev. B 73, 235411 (2006).

%Energy Band Gap Engineering of Graphene Nanoribbons
\bibitem{Han2007}
M. Y. Han, B. Oezyilmaz, Y. Zhang, and P. Kim, Phys. Rev. Lett, {\bf 98}, 206805 (2007).

%RKKY
\bibitem{RKKY}
M. A. Ruderman and C. Kittel, Phys. Rev. {\bf 96}, 99 (1954); T. Kasuya, Prog. Theor. Phys. {\bf 16}, 45 (1956); K. Yosida, Phys. Rev. {\bf 106}, 893 (1957); J. H. Van Vleck, Reviews of Modern Physics {\bf 34}, 681-686 (1962).

%Dielectric function and plasmon structure of stage-1 intercalated graphite
\bibitem{Shung1986}
K. W. -K. Shung, Phys. Rev. B {\bf 34}, 979 (1986).


%Indirect-exchange interactions in zero-gap semiconductors
\bibitem{Bastard1979}
G. Bastard and C. Lewiner, Phys. Rev. B 20, 4256  (1979).


%Dynamical polarization of graphene at finite doping
\bibitem{Wunsch2006}
B. Wunsch, T. Stauber, F. Sols, and F. Guinea, New Journal of Physics {\bf 8} (12),  318 (2006).


%RKKY in half-filled bipartite lattices: Graphene as an example
\bibitem{Saremi2007}
S. Saremi,  Phys. Rev. B {\bf 76}, 184430 (2007).


%Screening, Kohn anomaly, Friedel oscillation, and RKKY interaction in bilayer graphene
\bibitem{Hwang2008}
	E. H. Hwang and S. Das Sarma, Phys.\ Rev.\ Lett.\ \textbf{101}, 156802 (2008).


%Exchange interaction of magnetic impurities in graphene
\bibitem{Dugaev2006}
V. K. Dugaev, V. I. Litvinov, and J. Barnas,  Phys. Rev. B {\bf 74}, 224438 (2006).

%Screening effect and impurity scattering in monolayer graphene. 
\bibitem{Ando2006}
T. Ando, J. Phys. Soc. Japan {\bf 75}, 074716 (2006).

%\bibitem{Bunch2005}
%J. S. Bunch, Y. Yaish, M. Brink, K. Bolotin, and P. L. McEuen, Nano Letters {\bf 5}, 287 (2005).

\bibitem{Aristov1997}
%Indirect RKKY interaction in any dimensionality
D. N. Aristov, Phys. Rev. B 55, 8064 - 8066 (1997).

\bibitem{Litvinov1998}
%RKKY interaction in one- and two-dimensional electron gases
V. I. Litvinov and V. K. Dugaev, Phys. Rev. B 58, 3584 - 3585 (1998).

\bibitem{Balcerzak2007}
%A comparison of the RKKY interaction for the 2D and 3D systems and thin films
T. Balcerzak, JMMM {\bf 310}, 1651 (2007).

%RKKY exchange interaction within the parabolic quantum-well
\bibitem{Bak2001}
Z. B\c{a}k, Solid State Communications, {\bf 118}, 43 (2001)

%Two-impurity Anderson model for the transport through two quantum dots coupled via a quantum wire
\bibitem{Durganandini2006}
P. Durganandini, Phys. Rev. B, {\bf 74}, 155309 (2006)

\bibitem{Ziener2004}
%RKKY interaction in semiconductors: Effects of magnetic field and screening
C. H. Ziener, S. Glutsch, and F. Bechstedt, Phys. Rev. B 70, 075205 (2004).

%Anomalous properties of Ce-Cu- and Yb-Cu-based compounds
% Review
\bibitem{Bauer1991}
E. Bauer, J. Advances in Physics {\bf 40}, 417 (1991).

%RKKY interaction between Ce ions in CexLa1-xB6
%application of Coqblin-Schrieffer 
\bibitem{Schlottmann2000}
P. Schlottmann, Phys. Rev. B {\bf 62}, 10067 (2000).

%Magnetic exchange interactions in cerium compounds
%forth order Perturbation theory
\bibitem{Proetto1982}
C. Proetto and A. L\'{o}pez, Phys. Rev. B 25, {\bf 7037} (1982).


%Magnetic moment and valence fluctuations in a model for rare earth compounds
%forth order Perturbation theory
\bibitem{Hewson1977}
A. C. Hewson, J. Phys. C: Solid State Phys. {\bf 10} 4973 (1977).


%Crystal field effects on the magnetic susceptibility of mixed valence systems
%equation of motion
\bibitem{Mohanty1981}
J. C. Mohanty, P. K. Misra and S. D. Mahanti, J. Phys. C: Solid State Phys. {\bf 14} L1125 (1981).


%Effect of finite f level linewidth on the theory of the α-γ transition in Ce
%equation of motion
\bibitem{Alascio1973}
B. Alascio, A. L\'{o}pez and C. F. E. Olmedo, J. Phys. F: Met. Phys. {\bf 3} 1324 (1973).


\bibitem{BookMahan}
G. D. Mahan, {\it Many particle physics}, Plenum Press, New York (1991)


%Local moment physics in heavy electron systems
%REVIEW
\bibitem{Coleman2002}
P. Coleman, {\it Lectures on the Physics of Highly Correlated Electron Systems VI}, Editor F. Mancini, American Institute of Physics, New York (2002), p.79 - 160; cond-mat/0206003.

\bibitem{Proetto1981}
%Fourth-order effective Hamiltonian for the Anderson lattice
C. Proetto, A. Lopez, Phys. Rev. B 24, 3031 - 3036 (1981).



%-----------------------------
%dillute semiconductors
%-----------------------------


%Ferromagnetism in a dilute magnetic semiconductor:  Generalized RKKY interaction and spivn-wave excitations
\bibitem{Singh2003}
A. Singh, A. Datta, S. K. Das, and V. A. Singh, Phys. Rev. B 68, 235208 (2003).


%-----------------------------
%high tc superconductors
%-----------------------------


%Canonical perturbation theory and the two-band model for high-Tc superconductors
\bibitem{Zaanen1988}
J. Zaanen and A. M. Ole{\'s}, Phys. Rev. B {\bf 37}, 9423 (1988).




%Magnetic frustration in the three-band Anderson lattice model for high-temperature %superconductors
%only one schreiffer wolff transformation but in 4th order
\bibitem{Ihle1990}
D. Ihle and M. Kasner, Phys. Rev. B {\bf 42}, 4760 (1990).





%Apical Oxygen and Frustration Effects in CuO2 Lattice
% one fourth order SW transformation
\bibitem{Eid1995}
Kh. Eid, M. Matlak, J. Zieliski, physica status solidi (b), {\bf 187}, 589 (1995).


%-----------------------------
%variational wave study
%-----------------------------

%Two-impurity Anderson model: A variational study
\bibitem{Andreani2003}
L. C. Andreani and H. Beck, Phys. Rev. B 48, 7322 - 7337 (1993)

%Two Anderson Impurity problem: Kondo-doublets beyond the Kondo Limit
\bibitem{Simonin}
J. Simonin, {\it arXiv:cond-mat/0703531};%Kondo Quantum Dots and the Novel Kondo-doublet interaction
J. Simonin, {\it arXiv:cond-mat/0607620};
%Two Anderson impurities in the Kondo limit: A systematic study of the ground states of the many subspaces of the Hamiltonian
%variational wave function ???
J. Simonin, Phys. Rev. B {\bf 73}, 155102 (2006); J. Simonin, {\it cond-mat/0510580}.



%-----------------------------
%Perturbation Theory
%-----------------------------



%Diagrammatic approach to the intermediate-valence compounds
%forth order Perturbation theory
\bibitem{Grewe1981}
N. Grewe and H. Keiter, Phys. Rev. B {\bf 24}, 4420 (1981).


%On the absence of magnetic order in intermediate valence compounds
%forth order Perturbation theory
\bibitem{Kuramoto1981}
Y. Kuramoto, Z. Phys. B 40, 293 (1981).


%-----------------------------
%large n expansion
%-----------------------------

%Review of techniques in the large-N expansion for dilute magnetic alloys
%Large N expansion,  /not so necessary
\bibitem{Bickers1987}
N. E. Bickers, Rev. Mod. Phys. {\bf 59}, 845 (1987).

%-----------------------------
%bethe ansatz
%-----------------------------

%Analytical results for the Coqblin-Schrieffer model with generalized magnetic fields
% Bethe-Ansatz to Coqblin-Schrieffer
\bibitem{Bazhanov2003}
V. V. Bazhanov, S. L. Lukyanov, and A. M. Tsvelik, Phys. Rev. B 68, 094427 (2003).

%-----------------------------
%Nested Schrieffer-Wolff-Type
%-----------------------------


%Nested Schrieffer-Wolff-Type Transformations: Indirect Exchange Hamiltonian for a CuO2 Plane
\bibitem{Kolley1997}
E. Kolley, W. Kolley, and R. Tietz, phys. stat. sol. (B), {\bf 204}, 763 (1997).


%Ruderman–Kittel–Kasuya–Yosida interaction versus superexchange in a CuO2 plane in the limit U
\bibitem{Kolley1998}
E. Kolley, W. Kolley and R Tietz, J. Phys.: Condens. Matter {\bf 10}, 657 (1998).


\bibitem{Kolley1992}
%Fourth-order interactions in the canonically transformed d-p model for Cu-O superconductors
E. Kolley, W. Kolley and R. Tietz, J. Phys.: Condens. Matter {\bf 4}, 3517 (1992).


%Magnetic interactions in the Emery model
%two-stage schrieffer-wolff
\bibitem{Tien1994}
T. Minh-Tien, Physica C: Superconductivity {\bf 223}, 361 (1994).


%Generalized Schrieffer-Wolff Transformation of 2 Kondo Impurity Problem
\bibitem{Ong2006}
T. Tzen Ong and B. A. Jones, {\it arXiv:cond-mat/0602223}


%-----------------------------
%mixture perturbation and Schrieffer-Wolff-Type
%-----------------------------


%Exchange Interaction in Alloys with Cerium Impurities
\bibitem{Coqblin1969}
B. Coqblin and J. R. Schrieffer, Phys. Rev. {\bf 185}, 847 (1969).


%Nonequilibrium transport through parallel double quantum dots in the Kondo regime
\bibitem{Yang2005}
Y.F. Yang and K. Held,  Phys. Rev. B 72, 235308 (2005)

%Competing hybridization and consequences for magnetic ordering in ternary and quaternary correlated-electron systems
% couplin-schrieffer plus first fourth order term from SW
\bibitem{Castro1995}
C. Sanchez-Castro, B. R. Cooper, and K. S. Bedell, Phys. Rev. B {\bf 51}, 12506 (1995).


%Influence of the Crystalline Field on the Kondo Effect of Alloys and Compounds with Cerium Impurities
% more detailed derivation
\bibitem{Cornut1972}
B. Cornut and B. Coqblin, Phys. Rev. B {\bf 5}, 4541 (1972).


%Planar Coupling Mechanism Explaining Anomalous Magnetic Structures in Cerium and Actinide Intermetallics
%correction of Couplin-Schrieffer
\bibitem{Siemann1980}
R. Siemann and B. R. Cooper, Phys. Rev. Lett. {\bf 44}, 1015 (1980).


%Tunable Exchange Interaction in Quantum Dot Devices
\bibitem{Tamura2004}
H. Tamura, K. Shiraish, and H. Takayanagi,  Jpn. J. Appl. Phys. {\bf 43}, L691 (2004).

\bibitem{Schrieffer1966}
%Relation between the Anderson and Kondo Hamiltonians
J. R. Schrieffer and P. A. Wolff, Phys. Rev. {\bf 149}, 491 (1966).

\bibitem{BookFlensberg}
H. Bruus and K. Flensberg, {\it Many-Body Quantum Theory in Condensed Matter Physics: An Introduction}, Oxford University Press, Oxford (2004).

\bibitem{BookHewson}
A. C. Hewson, {\it The Kondo Problem to Heavy Fermions}, Cambridge University Press, Cambridge (2003).

\bibitem{note1} 
One could be temped to assume, that the sepatration of the interaction terms in Eq.~(\ref{eq:hamiltonian2}) resembles the distinction of different types of physical interactions. 
This qppears not to be the case.

%Diagonalisation of the generalised Anderson model
% Bethe-Ansatz to Coqblin-Schrieffer
\bibitem{Coleman1986}
P. Coleman and N. Andrei, J. Phys. C: Solid State Phys. {\bf 19} 3211 (1986).

%Spin coupling between magnetic impurities in metals
\bibitem{Zhang1986}
Q. Zhang and P. M. Levy, Phys. Rev. B {\bf 34}, 1884 (1986)

\bibitem{Brey2007}
%Elementary electronic excitations in graphene nanoribbons
L. Brey and H. A. Fertig, Phys. Rev. B 75, 125434 (2007).

\end{thebibliography}
\end{document}